  \magnification=\magstep1
 \settabs 18 \columns
    \voffset=0.20truein
\hsize=16truecm

\input epsf

\def\b{\bigskip}
\def\bb{\bigskip\bigskip}

\def\no{\noindent}
\def\r{\rightline}
\def\ce{\centerline}
\def\ve{\vfill\eject}

\def\r{\rightline}

\def\L{{\cal L}}

\def\harr#1#2{\smash{\mathop{\hbox to .25 in{\rightarrowfill}}
  \limits^{\scriptstyle#1}_{\scriptstyle#2}}}

\def\R{{\cal R}}

\def\today{\ifcase\month\or January\or February\or March\or April\or
May\or June\or July\or
August\or September\or October\or November\or  December\fi
\space\number\day, \number\year }

\r \today

\bb\bb\bb

\def\DD{\vec \bigtriangledown}


\def\w{\wedge}

\def\p{\partial}

\def\sqr#1#2{{\vcenter{\vbox{\hrule height.#2pt
\hbox{\vrule width.#2pt height#2pt \kern#2pt
\vrule width.#2pt}
\hrule height.#2pt}}}}

  \def\1/2{{\scriptstyle{1\over 2}}}
  \def\a/2{{\scriptstyle{3\over 2}}}
  \def\5/2{{\scriptstyle{5\over 2}}}
  \def\7/2{{\scriptstyle{7\over 2}}}
  \def\3/4{{\scriptstyle{3\over 4}}}

\font\steptwo=cmb10 scaled\magstep2


\def\picture #1 by #2 (#3){
  \vbox to #2{
    \hrule width #1 height 0pt depth 0pt
    \vfill
    \special{picture #3} 
    }
  }

\def\scaledpicture #1 by #2 (#3 scaled #4){{
  \dimen0=#1 \dimen1=#2
  \divide\dimen0 by 1000 \multiply\dimen0 by #4
  \divide\dimen1 by 1000 \multiply\dimen1 by #4
  \picture \dimen0 by \dimen1 (#3 scaled #4)}
  }

%
%

\def\sqr#1#2{{\vcenter{\vbox{\hrule height.#2pt
\hbox{\vrule width.#2pt height#2pt \kern#2pt
\vrule width.#2pt}
\hrule height.#2pt}}}}

\def \r{\rightarrow}

\b

\vskip-1cm

\ce{\steptwo Action Principle for Hydrodynamics and Thermodynamics}

\ce{\steptwo including general,  rotational flows}
\bb

\ce{\it C. Fronsdal}
\b

\ce{\it Depart. of Physics and Astronomy, University of California Los Angeles
90095-1547 USA}
 
\bb

\no {\it ABSTRACT}\quad

This paper presents an action principle for hydrodynamics and thermodynamics of fluids that includes general, rotational flows, responding to a challenge that is more than 100 years old. It has been lifted to the relativistic context and it can  be used to provide a suitable source of rotating matter for Einstein's equation.

The theory is a combination of Eulerian and Lagrangian hydrodynamics, with an extension to thermodynamics. In the first place it is an action principle for adiabatic systems, incorporating the fundamental conservation laws of hydrodynamics as well as the Gibbsean variational principle of equilibrium thermodynamics. But it also provides a framework within which dissipation can be introduced in the usual way, by adding a viscosity term to the momentum equation, one of the Euler-Lagrange equations. It is a development of  the Navier-Stokes-Fourier approach, the principal advantage being a greatly expanded predictive power. The approach is traditional in so far as it is based on the same conservation laws, with the difference that those conservation laws are united in 
an action principle. The theory  has a Hamiltonian structure with a natural concept of energy, as a first integral of the motion, conserved (in the absence of viscosity) by virtue of the Euler-Lagrange equations. 

Two velocity fields are needed,
one the gradient of a scalar potential, the other the time derivative of a vector field (vector potential). Variation of the scalar potential gives the equation of continuity and variation of the vector potential yields the momentum equation into which the viscosity can be introduced.   The nature of the velocity field(s) has a profound effect on the structure of the energy momentum tensor which, in turn, affects the  coupling to the gravitational field, as is discussed elsewhere (Fronsdal 2015). The concepts are intimately related to work by Lund and Regge and others on vortex strings.

\b

This paper is dedicated to the memory of my friend Volodya Kadyshevsky.

 \ve
 
\no{\bf I. Introduction}

Many branches of theoretical physics have found their most powerful formulation as action principles. What most distinguishes these theories  ({\it e.g.} QED) is their
superior predictive power, providing values of experimentally verifiable properties to as much as 10 significant figures. The prediction of masses of particles before their actual discovery (the heavy mesons predicted by the standard model)  is without precedent. In contrast, branches of physics that have not yet been cast in the form of action principles (e.g. hydrodynamics) 
do not, for that very reason, have enough constraints placed on them. They allow a great amount of freedom that makes it too easy to describe observed phenomena
and, as a result,  their predictive power is limited. The Navier-Stokes equation is useful, 
and a valid expression of physical principles, but its most common employment has 
been to measure viscosities.

An action principle is known for Eulerian hydrodynamics; it is limited to potential flows and its use is restricted (Fetter and Walecka 1980).   But it has an extension that includes adiabatic thermodynamics and applications are plentiful if not widely known; some will be quoted in this paper. Some of these applications give clear
evidence for the increase in predictive power that can be gained by adopting an action principle formulation. The alternative `Lagrangian' version of hydrodynamics also has an action principle
formulation, but it does not incorporate an equation of continuity.
This paper presents a general action principle for hydrodynamics,
with an extension to thermodynamics, that includes the equation of continuity and that - the main novelty - encompasses general velocity fields including vorticity.

  The Navier-Stokes equation has maintained a dominant position in hydrodynamics for more than 100 years. It has known numerous successes and no real failures, but see Brenner (2013) and especially the review by Martin (2010). 
It suffers, nevertheless, from  an ambiguous relationship with the energy concept. 
This is understandable, since it was designed to describe processes in which energy is dissipated, but it implies a lack of completeness for which most workers have felt a need to  compensate. It is  usually supplemented by an `energy equation'.
 But the choice of an expression for this `energy' is seldom canonical 
and cannot be fully justified. Dissatisfaction with this approach is occasionally expressed in the literature, as in this example  (Khalatnikov (1965). After listing a
  ``complete system of hydrodynamic equations" (equations 2-5) he presents 
one more,   ``the energy conservation law $\p E/\p t + {\rm div} Q = 0$", and then he says: ``It is necessary to choose the unknown terms in Eq.s (2-5) in such a way that 
this last equation be automatically satisfied."

This paper does not claim to have discovered the future theory of hydrodynamics. It merely claims to prove that an action principle exists that embraces a major part of the standard theory,  and it gives some
hints to support the contention that it has a greater power of prediction than the 
standard approach. It is especially powerful in dealing with mixtures.

 \b

\ce{\bf Degrees of freedom}

To formulate an action principle one begins by choosing the dynamical variables.
The principal variable is a velocity field but, in order that Euler-Lagrange equations
give a differential equation for the velocity,  the velocity has to be represented as a derivative of a more basic variable. In the case of potential flows it is a space derivative,  $\vec v = -\DD \Phi$, the alternative is a time derivative, $\vec u = d\vec X/dt$.  Both are basic in studies of turbulence.

The study of hydrodynamics begins by counting the number of degrees of freedom,
the density and the flow velocity together makes four of them. The restriction to potential flow has only two, the density and the velocity potential make one canonical pair.

Minkowski (1908) and Tolman (1934) promoted the 3-vector velocity field to a relativistic 4-vector field; the former in electrodynamics and the latter in General Relativity. Tolman's 4-velocity is normalized to reduce the number of degrees of freedom to 3 (plus the density), but neither Minkowski nor Tolman introduced a canonical structure and no equation of continuity appears.

Schutz (1967) attempted to construct an action principle for relativistic thermodynamics in which the principal variable 
would be a 4-velocity. To construct an action principle it was necessary that this 4-vector 
field be a derivative of something more basic. The gradient of a scalar field gives only
one degree of freedom, so Schutz added other 4-vector fields derived from additional tensor fields. This theory has not known further development. It appears to have too many independent degrees of freedom (more than 8).

The theory proposed  here  has 3 velocity degrees of freedom, as preferred. The density is a fourth independent variable and the four together make 2 canonical pairs.
Because we are aiming at a relativistic theory another way to count degrees of freedom is more to the point, and easier. 
The relativistic generalization of the model developed here will be presented in another paper, except for a very brief sketch. But it is relevant to point out that the 4 degrees of freedom that appear in traditional non relativistic hydrodynamics  are faithfully reproduced.
The inclusion of a scalar velocity potential is inevitable in view of the need to generate the equation of continuity, a scalar equation, as one of the Euler-Lagrange equations of motion. 
The additional velocity field introduced here is related to a relativistic 2-form of the  type introduced by Ogievetsky and Palubarinov (1964) - a gauge theory. It is well
known that it has one propagating mode; hence one canonical pair of independent physical variables.
\b

\ce{\bf Two kinds of flow}
In the case of potential flows of a homogeneous fluid the velocity field is the gradient of a scalar potential. The Eulerian point of view is natural and convenient; the density and the scalar potential form a canonically conjugate pair of dynamical variables and variation of the scalar velocity potential yields the equation 
of continuity.  In general, the variational 
approach requires a vector potential as well, such as appears in the alternative, Lagrangian formulation, the velocity is represented as the time derivative $\dot{\vec X}$ of a vector field (a vector potential). But the density does not have a conjugate momentum and there is no equation of continuity. 
 
It is a well known and remarkable fact that some phenomena in fluid mechanics 
are characterized by two kinds of flow. The vortices that are seen in the wakes of 
ships are created locally; the angular momentum decreases as the inverse square of the distance from the center; this is characteristic of potential flow. In experiments involving fluids  confined between a pair of concentric, rotating cylinders (Couette flow) it is observed that motion generated by steady rotation of the enveloping vessel tends to create a very different type of flow pattern that is close to that of a solid body, accompanied by local vortices that are described in terms  of potential flow. 

Here the Navier-Stokes equation comes through with the successful prediction that, in the presence of viscosity, precisely these two kinds of stationary flow are possible.

In their study of flow patterns in super conducting Helium, Landau (1954) and  Feynman
(1954) describe the small vortices in terms of potential flow (`phonons'), while the 
alternative, competing flow is referred to as `motion of the system as a whole';
or solid-body flow (`rotons'). This point of view is still dominant in the more recent literature, see for example the review by Fetter (2009). No independent field theoretical degree of freedom was associated with the second kind of flow, in this context.
But in some important studies of vortex lines the flow vector $\dot{\vec X}$ is a prominent, dynamical field. In an 
 approach initiated by Onsager and pursued in two remarkable papers, by
Rasetti and Regge (1975) and  by Lund and Regge (1976), they do introduce a separate  velocity field, associated with the motion of vortex lines. These papers are physically and
mathematically related to the present paper.

The variational principle for hydrodynamics that is presented here combines potential flows and `solid-body' flows, just as Navier-Stokes does, but two independent vector fields are needed, one derived from a scalar potential, the other from a vector potential.

The introduction of a pair of velocity fields is by no means new. See the quoted review by Martin (2010) as well as earlier reviews quoted therein.


\b

\ce{\bf Thermodynamics} 

The emphasis is on hydrodynamical concepts, but we wish to advocate
the broader perspective of thermodynamics, where applications of the Navier-Stokes equation have been less rewarding.  
Fig. 1 is meant to evoke the structure of the Gibbsean geometric concept of 
thermodynamics, the imbedding space of unconstrained dynamical variables over which the variation is to be carried out and the subspace of extrema that are the physical states. The variation is carried out with the entropy distribution fixed,
the choice of entropy determines a well defined Lagrangian and corresponds to  one of the subspaces illustrated by the blue lines (approximately parabolas) in the figure.
The red line (at the bottom, nearly straight) is the collection of equilibrium states,
one for each adiabatic system. `Non-equilibrium thermodynamics' refers to the problematics that must be faced when only the states of equilibrium  are known. We take the position that adiabatic thermodynamics is a developed discipline and that this gives a much wider platform from
which to launch a study of dissipation. A change of state that involves viscosity and other forms of dissipation is best understood when it can be described as a quasi-static evolution along a sequence of adiabatic equilibria; it seems evident that such studies must be built on a coherent account of the adiabatic systems themselves, not just their equilibria (Prigogine 1965). An adiabatic system gives a
meaning to the concept of equilibrium, a family of adiabatic systems is  needed to
define dissipation.

\b

\epsfxsize.5\hsize
\centerline{\epsfbox{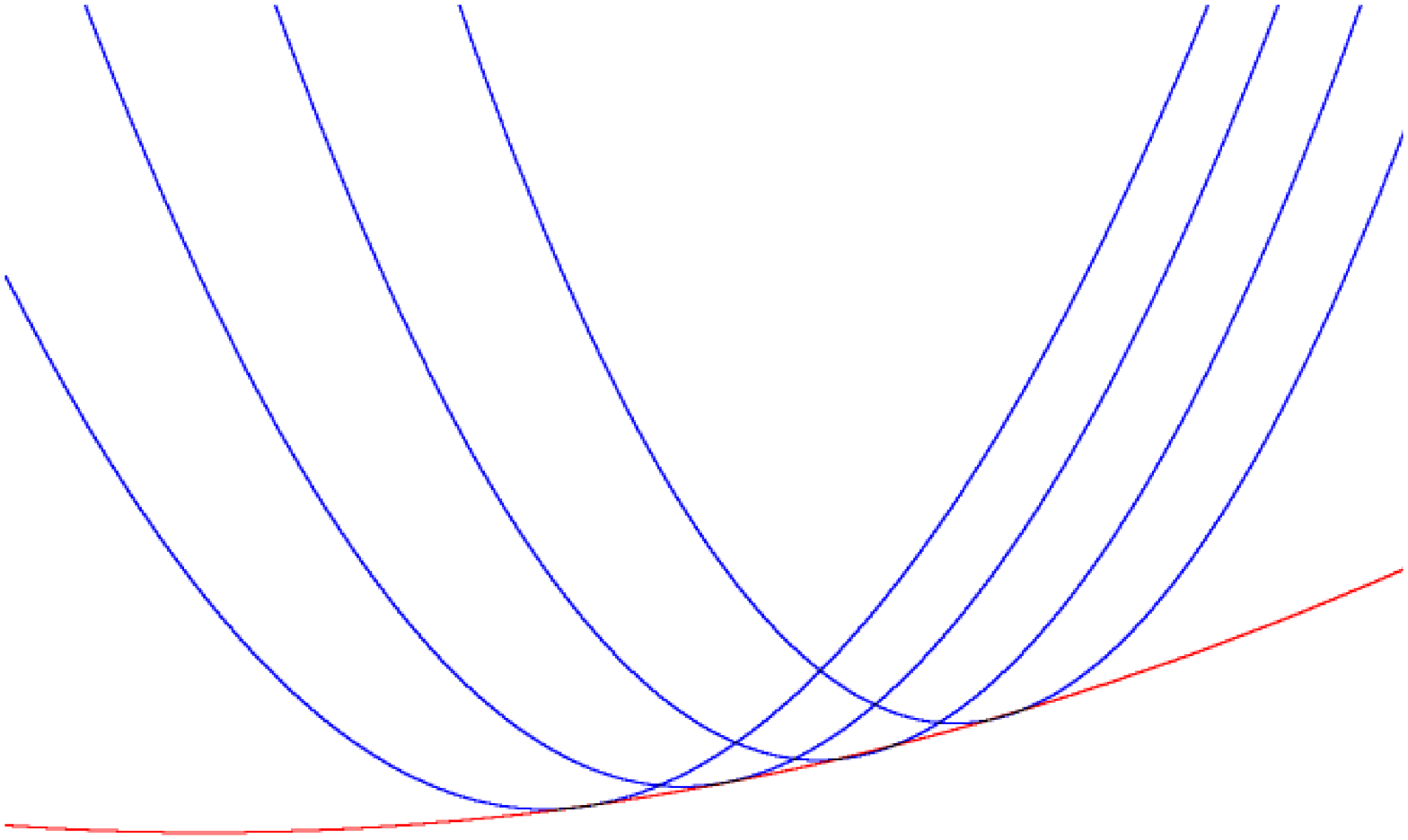}}
\vskip0cm
 
 Fig.1. Geometric presentation of Gibbs' thermodynamics.
\bb


The dominant strategy that has been pursued under the banner of nonequilibrium thermodynamics  is based on a collection of conservation laws, subsequently modified to account for dissipation. The most important are the equation of continuity (mass conservation) and the Navier-Stokes equation (in the absence of viscosity: momentum 
conservation). The Navier-Stokes equation has a long and impressive history, including
numerous successful applications. As we look for a variational principle we must try to incorporate as many properties of Navier-Stokes as we can; that is the strategy that has
characterized this work.

The original aim was  to find appropriate matter sources, including rotational motion, for the dynamical metric field of General Relativity (Fronsdal 2007, 2015) and the present work represents a step in that direction.  This report remains within the non relativistic context, classical hydrodynamics and thermodynamics. Nevertheless one basic feature of the relativistic extension must be pointed out here.

The essential step that lifts classical hydrodynamics to
the relativistic context is commonly regarded  to be the promotion of the 3-vector velocity to a timelike, 4-vector velocity field.
This is natural in the case of a gradient vector field but not when the velocity is presented as a time derivative of a 3-vector field. In that case the relativistic velocity is the dual of an exact 3-form. This make a very big difference in the
structure of the energy momentum tensor; see the last paragraph of this paper.   Implications for General Relativity are described in a recent paper (Fronsdal 2015).
Applications to electromagnetism in fluids have  been initiated.

\b
 
\ce{\bf Summary of the paper}

Section II presents the well known variational formulation of potential flows in
hydrodynamics, and a short account of the unification with thermodynamics. The next section describes the strategy that was chosen to look for an action principle of wider generality: to focus on the special case of cylindrical Couette flow,  with an unusual emphasis on laminar flow,  in this simplest context in which non-potential flows impose themselves as inevitable.  Section IV reviews the classical 
application of the Navier-Stokes equation to this system and contrasts it with the failure of potential theory to deal with it, to discover the tight spot and the remedy. 

Section V presents the proposed solution to the problem, a solution that is both simple and natural. It combines Eulerian and  `Lagrangian' hydrodynamics to describe both potential and rotational flows. The extension that includes viscosity is natural and standard, Section VI.

Section  VII is a discussion of the relative merits of the variational principle and the Navier-Stokes equation in the context of adiabatic dynamics where both are applicable.
The main issue is the role of energy. Once a system of conservation laws is obtained the extension that includes 
viscosity is natural and standard. It is suggested that the inclusion of two types of velocity fields may affect stability studies.    This section is followed by a brief study of plane Couette flow, where the relationship between the two theories turns out to be very different. The variational approach has the greater predictive power, in this case, but experimental confirmation must wait for more accurate temperature measurements.

Section IX is a very brief summary of  main conclusions and  possible applications.
    
\bb

\no {\bf II. The Fetter-Walecka action principle}

A variational formulation of simple hydrodynamics may be found in a book by Fetter and Walecka  (1980). The action is
$$
A= \int dt \int d^3 x \,\L, ~~~\L =  \rho(\dot\Phi - \vec {\vec v}\,^2/2) -W(\rho),\eqno(2.1)
$$
with the definition 
$$
\vec v = - {\rm grad}\,\Phi.\eqno(2.2)
$$
That is a strong restriction on the velocity field.

The equations of motion are the Euler-Lagrange equations derived by variation of the two scalar fields. Variation of $\Phi$ gives the
equation of  continuity,
$$
\dot\rho+\DD \cdot(\rho\vec v)=0\eqno(2.3)
$$
and variation of $\rho$,  
$$
\dot \Phi -  {\vec v}\,^2/2 = {\p W\over\p \rho},\eqno(2.4)
$$
(A Lagrange multiplier is included in the field $\dot\Phi.$)~ or
$$
\rho\dot{\vec v} +\rho \DD{\vec v}^2/2=-\DD \,p,~~ p := \rho{\p W\over \p\rho}-W,\eqno(2.5)
$$
where $p$  is the pressure. The two terms on the left side can be combined, 
$$
 \dot{\vec v} +\DD{\vec v}^2/2={D\vec v\over  Dt} :=\dot{\vec v} + (\vec v\cdot \DD)\,\vec v.\eqno (2.6)
$$
That the substantial derivative $D\vec v/Dt$ appears here comes
about because of the identity
$$
\DD\, {\vec v}\,^2/2 = (\vec v\cdot \DD)\,\vec v,
$$
which is true by virtue of  (2.2). 
Eq. (2.5) is the Bernoulli equation, written in the form in which it is most easily compared to the Navier-Stokes equation. It is the gradient of the `integrated Bernoulli equation' (2.4).

To gain some impression of the significance of this formulation see the review by Fetter (2009), where the quantization of vortices (Onsager, see Eyink and Sreenivasen 2006)  is expressed in terms of the one-valuedness of exp$(i\Phi/\hbar$ (constant factors omitted). 

The Bernoulli equation is a mainstay of hydrodynamics and its use is fully justified by its success. But when it is done within the context of a thermodynamical system
it turns out that it can derived from more fundamental axioms, under certain conditions.
\b

\ce{\bf Thermodynamics}

This action principle can be combined with Gibbs' thermodynamic principle of minimum energy (Gibbs 1878). Applied to a one component thermodynamic system the Lagrangian density takes the form
$$
\L =  \rho(\dot\Phi - \vec {\vec v}\,^2/2) -f(\rho,T) - sT.\eqno(2.7)
$$
Here $f$ is the free energy density and $s = \rho S$ is the entropy density.
Variation of $T$ (the temperature) gives the adiabatic equation that permits the 
elimination of the temperature in favor of the specific entropy density $S$, assumed uniform. Variation of  $\Phi$
gives the equation of continuity and variation of the density gives
$$
\dot \Phi -  {\vec v}\,^2/2 = \mu,\eqno(2.8)
$$
where $\mu$ is the chemical potential. The chemical potential can be expressed in terms of the pressure and the entropy, which leads to the Bernoulli equation or, in the static case, to the hydrostatic equation. But this requires knowledge of the entropy. In the case of an ideal gas  $\mu$ can expressed in terms of the temperature and we obtain in that case
$$
\dot \Phi -  {\vec v}\,^2/2 = C_VT.\eqno(2.9)
$$
This form of the equation is, in our opinion, much to be preferred, for it requires no knowledge of the entropy. The comparison of theory and experiments and, consequently, the interpretation,  would be greatly facilitated if it were possible to make more accurate measurements of the temperature profile.

It is of great interest to understand the position of the Bernoulli equation in this
thermodynamic context. Variation of the action with respect to the density gives
$$
\dot\Phi - \vec v^2/2 = {\p \over \p \rho}(f+sT).
$$
Taking the gradient gives an equation that resembles the Bernoulli equation, but to obtain the familiar form in terms of the pressure it is necessary to make two
assumptions: 1. that the entropy density $s$ is a linear function of the density, $s = \rho S$, and 2. that the specific entropy $S$ is uniform. Hence the remarkable conclusion that every successful application of the Bernoulli equation implies that the specific entropy is uniform. This is widely known (Stanyukovich 1960) but it  is seldom pointed out.

Up to this point in our work (Fronsdal 2014a, 2015) the attention has been focused, out of necessity, on the special case when the flow velocity is a gradient, the curl being zero,
$$
\vec v = -  {\DD} \Phi~~~\rightarrow ~~~ {\DD} \wedge \vec v = 0.
$$
Although  restrictive, this limitation was accepted as an apparently necessary condition to formulate hydrodynamics (and thermodynamiccs) as a Laxgrangian field theory.  Many problems in thermodynamics and hydrodynamics involve  
no flow and in many others the flow is irrotational. 
But a relativistic Lagrangian is needed for General Relativity and this is one of many applications where the restriction to potential flows must be overcome.  To discover how to achieve greater  generality we shall now turn our attention  to a simple 
system where a generalization is obviously and urgently needed. 
 
\epsfxsize.5\hsize
\centerline{\epsfbox{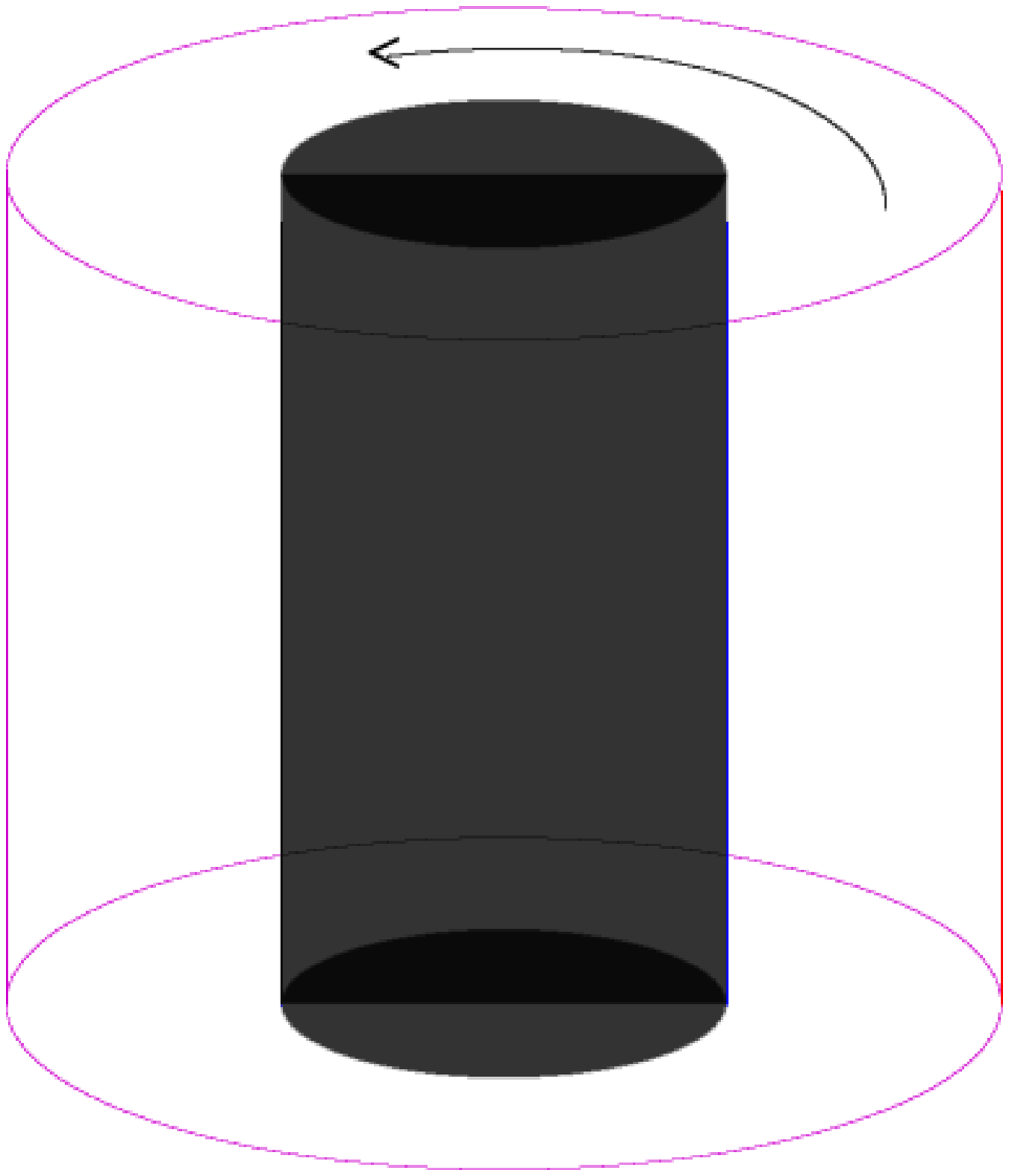}}
\vskip0cm

   Fig. 2. Cylindrical Couette flow is the steady, horizontal, rotational flow between two concentric cylinders.
\b   

\no{\bf III. Couette flow, potential flow}

We shall consider a situation that is effectively 2-dimensional because of translational symmetry, when nothing depends on a vertical coordinate $z$ and the flow is parallell to the horizontal $x,y$ plane. Couette flow is the flow of a fluid between a pair of concentric cylinders that can be rotated around a common (vertical)  axis.
With both cylinders at rest we postulate an initial state in which the space
bounded by the two cylinders is filled with a fluid at rest, with all variables
time independent, and uniform. The effect of gravity will be neglected. The cylinders are long enough that end effects can be neglected as well.\footnote*{
According to a Merriam Webster web page,  the term Couette flow is derived from ``French couette, machine bearing, literally, feather bed, from Old French coute, cuilte quilt, mattress." To make up for this slight we cite several of Couette's pioneering papers (Couette 1887,1888,1889,1890).}

We begin to rotate the inner cylinder. For this to have any effect on the fluid
we need to postulate a degree of adherence of the liquid to the surface of the inner cylinder.  We suppose that a state of stationary rotation is approached asymptotically and that there is no loss of energy at the wall in the limit. This is summed up by the no-slip  boundary condition
$$
\vec v|_{r=r_0} =\vec v|_{\rm inner~boundary} = \omega_0(-y,x,0).
$$
The angular velocity $\omega_0$ is  that of the cylinder.

The no-skip boundary condition has been widely applied, and because we wish to
compare our approach to the traditional one it serves our purpose to do the same.
See Brenner (2011), Priezjev and Troian (2005), Dukowicz, Stephen,  Price and  Lipscomb (2010),  Goldstein, Handler and Sirovich (1993).
  
The rotational axis is the $z$ axis. The coordinates are inertial and Cartesian.   No boundary conditions are imposed on the velocity at the outer wall, so far.

We propose to treat this system, with fixed boundary conditions, as an adiabatic system, and the stationary solution as its equilibrium configuration. We begin with the Fetter-Walecka Lagrangian density
$$
\L= \rho(\dot\Phi - {\vec v}^2/2) - f-sT.\eqno(3.1)
$$
The problem is essentially 2-dimensional and
$r$ is the cylindrical, or 2-dimensional-radial coordinate. 
When all the variables are time independent the Euler-Lagrange equations reduce to
$$
{\rm div}(\rho\vec v) = 0,~~~\dot\Phi - {\vec v}^2/2 = \mu,\eqno(3.2)
$$ 
where $\mu$ is the chemical potential and $\dot\Phi$ is a constant. In the case of an ideal gas ,
$$
\dot\Phi - {\vec v}^2/2  = C_V T.\eqno(3.3)
$$
(See Eq.s(2.8) and (2.9).) If we assume that the specific entropy is uniform,  then this last equation can be transformed to the more familiar Bernoulli equation,
$$
\rho\dot{\vec v}+\rho\DD ({\vec v}^2/2) =- \DD p.\eqno(3.4)
$$

For this stationary state the flow lines are circles, $r^2 = x^2 +y^2 = $ constant,
$$
\vec v= \omega(r)(-y,x,0).
$$
It is a field with vanishing curl only if
$$
\omega(r) =  a r^{-2},~~~ a = {\rm constant},
$$
and even then it is not, in the strict sense, a gradient. Although
$$
\vec v =  a \,\DD\,\theta,~~~ \theta := \arctan {y\over x},\eqno(3.5)
$$
the scalar $\theta$ is not one-valued, though the vector field is. Thus, by allowing the velocity potential to be multivalued, 
$$
\vec v= {a\over r^2}(-y,x,0) = -\DD\Phi,~~~\Phi = a\theta,~~~ a = {\rm constant},
$$
one  extends the approach to include special rotations, with $\DD \wedge \vec v 
=0$ away from the origin  (though ill defined at that one point). In the present case the origin is outside the vessel and the curl is zero at all points of the fluid.
It is customary and convenient to encompass this situation in the concept ``potential flow".

This is the only possible horizontal,  stationary, potential flow. The angular
velocity is greater at smaller radius; it is therefore unsurprising that it is driven by 
rotating the inner cylinder and that it settles according to the no slip condition.
There is no opportunity to adjust the outer boundary conditions; either the fluid
slips there or else the angular speed of the outer cylinder must be adjusted in accordance with $\omega(r)r^2 = a.$  

The divergence of $\vec v$ (for any choice of $\omega(r)$) is zero, so the equation of continuity reduces to $\vec v \cdot \DD \rho = 0$, 
requiring $\rho$ to depend on  $r$ and $z$ only.  From now on it is taken for granted that $\rho$ is a function of $r$ only.

This type of stationary motion has been observed, see for example, Joseph and Renardy
(1985), but the principal goal of most experiments has been to measure the onset of turbulence as, with increasing rotational speeds, the laminar flow breaks down
(Couette 1887).  Our present interest is focused, instead, on laminar flow at low angular speeds.

The principal equation of motion, Eq. (3.3), reduces, in the case that the velocity takes the form (3.5), in the case of an ideal gas, to
$$
{a^2\over 2r^2} = C- C_VT,~~~C = {\rm constant}.\eqno(3.6) 
$$
It appears that this  temperature lapse has not been measured. The experiment would be difficult to interpret because of the suspected heat flow caused by the friction,
within the liquid and between the liquid and the wall.    
The equation is consistent with the interpretation of the fluid as a  collection of classical particles, and in full agreement with the Navier-Stokes equation. \footnote*{ It is common, in the case of liquids, to treat them as incompressible. 
In this case it is justified to treat the pressure gradient as unknown, to be determined by the need to get reasonable solutions to the equations.
But of course this puts severe limits on the field of applications and on the  predictive power of the theory.}  
As is seen from Eq. (3.4), the kinetic energy acts as an effective potential and  gives rise to a 
repulsive,  radial force, balanced by  the negative of the pressure gradient.

\epsfxsize.5\hsize
\centerline{\epsfbox{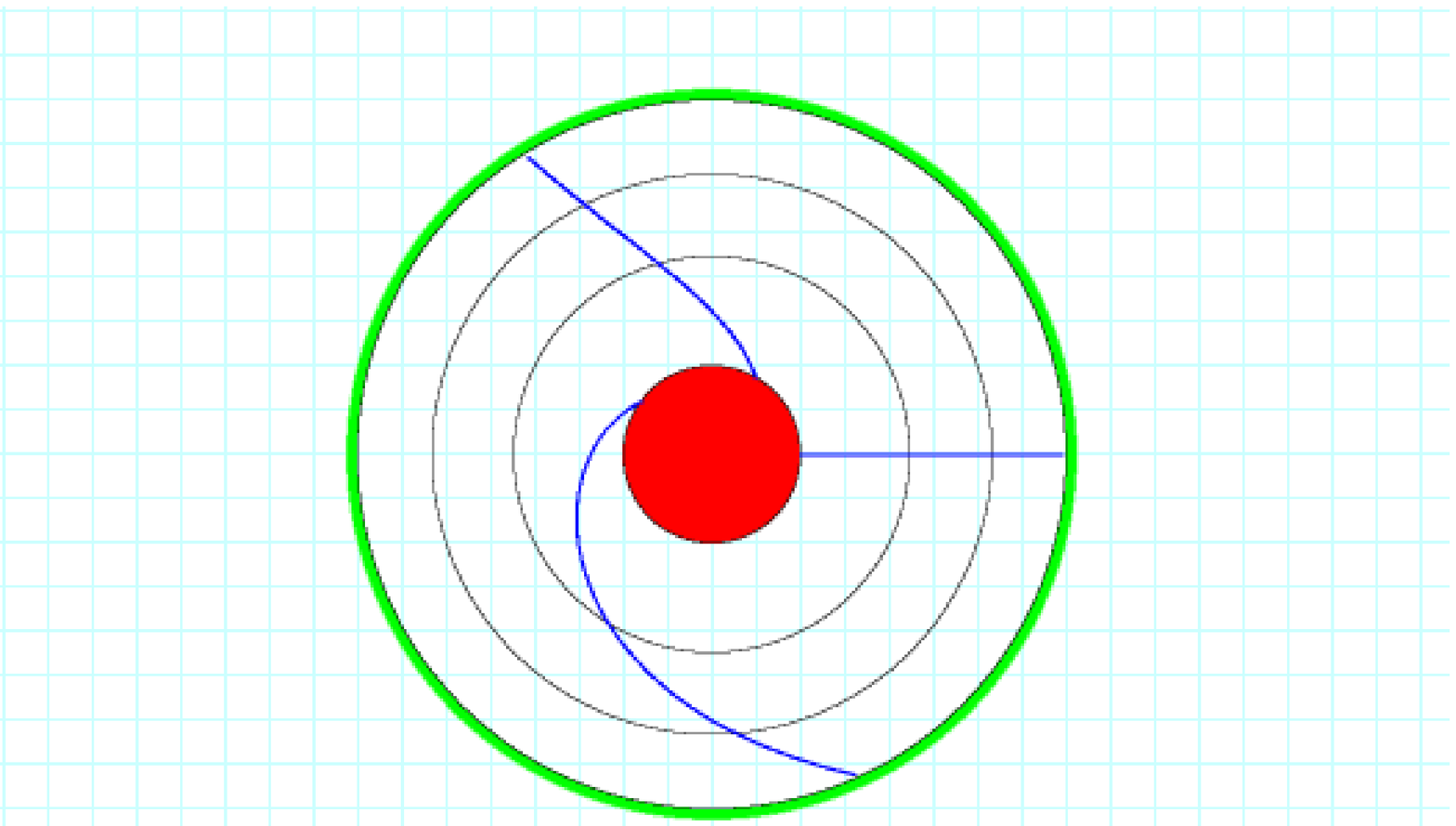}}
\vskip0cm

Fig. 3. Snap shot of Couette flow. The horizontal line shows the position of a set of particles at $t = 0$; the two curved lines the same particles at later times. The inner cylinder is rotating clockwise, dragging the fluids along with it.
  \bb

\ce{\bf`Solid-body' flow}

Consider next the complimentary experiment in which there is friction at the outer cylinder only, making it the driver of the motion. The problem is the same, except for different boundary conditions,
$$
\vec v|_{r=r_1} =\vec v|_{\rm outer~boundary} = \omega_1(-y,x).
$$
Potential motion provides only one solution, and here its application is anti-intuitive since it has the angular velocity increasing towards the center, away from the driving wall. For the general problem, in which the two cylinders move independently, we have two independent boundary conditions but a one dimensional space of potential flows. 
Clearly we have come to a situation where we cannot limit our attention to gradient velocity fields. Still,  this is not the end of the action principle approach to hydrodynamics. 

It is reported that the motion of a liquid, even a rarified gas, driven by the rotation of the outer cylinder, tends towards the motion of a solid body. See for example 
Andereck, Liu and Swinney (1986), de Socio,  Ianiro and Marino (2000).
As the speed is increased, various instabilities set in, but at present our only concern is to understand the laminar motion observed at low speeds.

The motion of a rigid body is characterized by the velocity field
$$
\vec v = b (-y,x,0),~~~b ~{\rm constant},
$$
and this is not a gradient. If we analyze this flow in the same manner
that we did  potential flow; that is, if we simply use the above velocity
field in the field equations, then we obtain similar results; in particular, instead of (3.6),
$$
{b^2\over 2}r^2 = C- C_V T.
$$
But note that to get this unphysical result we applied the equations of the potential theory beyond their domain of validity. Here the left side has the wrong sign,  the force that it implies is attractive (towards the center). This mesmerizing error in sign was a problem that frustrated all efforts, until the only possible solution popped into view. 
 
We have to escape from under the restriction to gradient velocities
and  a generalization of the action principle is required.  It is natural to ask how  
Navier-Stokes handles handles the situation. 

\bb

\no{\bf IV. Navier-Stokes equation}

  The standard treatment of non-potential flows 
is based on the continuity equation and the  Navier-Stokes equation (Navier 1827, Stokes 1843, Navier 1882),
$$
\dot\rho+{\rm div}(\rho\vec v) = 0, 
$$
$$
\rho\Big(\dot{\vec v} + (\vec v\cdot \DD) \vec v\Big)= - \DD p - \mu \Delta  \vec v.\eqno(4.1)
$$
This allows for
flows of both kinds, potential flow and solid body flow. The new elements are three. First of course, the nature of the velocity field is not constrained to be a gradient. In the second place the
term $\rho \DD {\vec v}^2/2$ in (3.4) has been replaced by the term 
$\rho(\vec v\cdot \DD) \vec v $ in (4.1). Finally,  there is  
the viscosity term $\mu \Delta \vec v$.  

If the coefficient $\mu$  (the dynamical viscosity) vanishes,   of the second equation only the radial component
remains. It is an ordinary differential equation for the functions $\omega,\rho$ and $p$ and the system  is under-determined. If $\mu \neq 0$, the tangential projection
 imposes the additional requirement that
$$
\Delta \vec v = 0.\eqno(4.2)
$$ 
This leads to unique solutions for reasonable boundary conditions, either or both cylinders driving. The fact that uniqueness is obtained only when the viscosity is taken into account may seem a little odd,
 since there is no lower limit on the value of the coefficient $\mu $.  Yet this may
 seem natural since we are considering stationary solutions only; the stationary  solutions  are those that avoid the dissipation induced by viscosity. And this is the standard treatment, so we proceed, with a mental reservation. See Section IX.
 
  Given the form
$$
\vec v= \omega(r)(-y,x,0)
$$
of the velocity field (no longer required to be a gradient) within the class of flows under consideration, the general solution of Eq.(4.2)  is
$$
\vec  v= \omega(r)(-y,x,0),~~~\omega(r)  =  {a\over  r^2} + b,\eqno(4.3)
$$
$a$ and $b$ constants; the two types of flow already considered are the only ones allowed by Eq. (4.2). 

The boundary conditions at $ = r_0, r_1$ give us
$$
\omega(r_0) = a{r_0}^{-2} + b = \omega_0,~~~
\omega(r_1) =  a {r_1}^{-2}  + b =   \omega_1,
$$
When the $`a'$ term dominates we have the highest angular velocity at the inner surface; this is as expected when the inner cylinder is driving. If the $`b'$
term dominates we have nearly constant angular velocity, as in the case of a solid body.   If both cylinders are rotating, in opposite directions, and both are driving, then $\omega(r)$ will have a change of sign. Explicitly,
$$
 a= {\omega_0-\omega_1\over r_0^{-2}-r_1^{-2}},
$$
and
$$
b= {1\over r_1^2r_2^2}{r_1^2\omega_1-r_0^2\omega_0\over r_0^{-2}-r_1^{-2}}.
$$

The result is that Navier-Stokes, with non zero viscosity, has just one extra solution besides the gradient, allowing it to satisfy no slip boundary conditions for all values of the 
angular velocities of the two cylinders.  This additional solution, with  $\vec v \propto (-y,x,0)$, is the same as the static state 
observed from a rotating reference frame; its existence is required by the relativistic equivalence theorem. And the fact that the Navier-Stokes equation singles out just two, 
radically different kinds of flow, and nothing else, is highly
significant.

The salient differences between Navier-Stokes and the theory attempted in the 
preceding subsection are the following. As noted, the term $\DD v^2/2$ in
Eq.(3.3) is replaced by the term $\vec v\cdot\DD \vec v$ in (4.1).
One verifies that, when the velocity field is a gradient, both expressions are equal, but in the case of solid body rotation there is a change of sign,
$$
\DD{\vec v}^2/2 = b^2\vec r~~{\rm but} ~~\vec v\cdot\DD \vec v = -b^2\vec r.
$$
We need to understand what lies behind this change of sign.

\vskip1cm

 \ce{\bf The centrifugal force}

In particle physics the dynamical variable is the position of the particle.
The `fictitious'  centrifugal force in particle physics can be seen as coming from an effective `kinematic' potential. The force is $\omega^2 \vec r $ directed outwards,
and this is $-\DD[-{\omega^2}r^2/2] $, so the effective potential is the negative of the kinetic energy. The unexpected sign comes from the fact that the the origin of this potential is in the term $ m\dot {\vec x}^2/2 $ in the Lagrangian; this term appears with the same sign in the Hamiltonian.   Let us say it in another way, the term in question appears with  positive sign in the Lagrangian and with the same sign in the Hamiltonian; but its contribution to the equation of motion
is opposite in sign from that of a normal potential. We have seen that our potential model when used outside its domain of validity, gives the wrong sign in the equation of motion.

\b

\epsfxsize.5\hsize
\centerline{\epsfbox{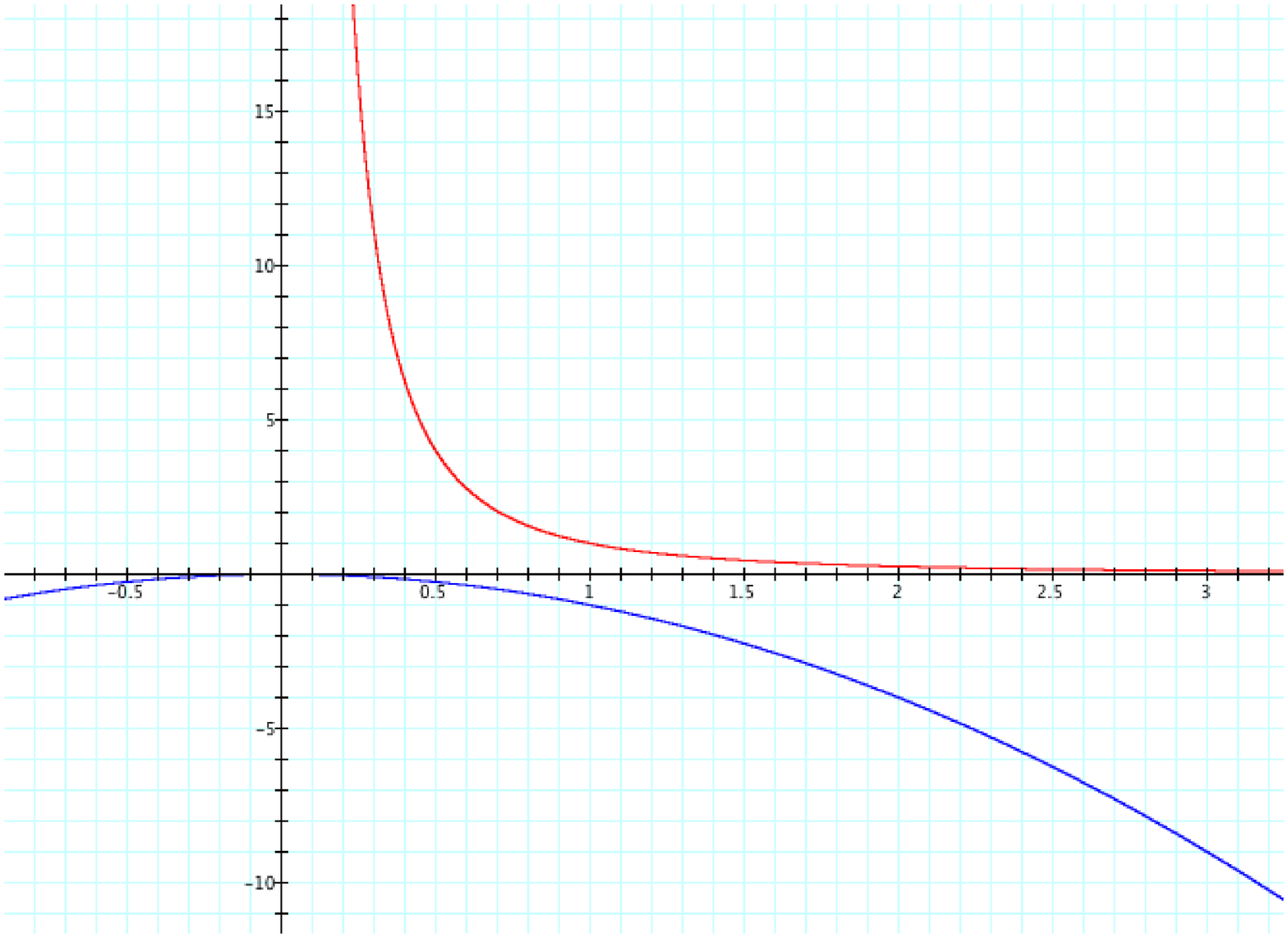}}
\vskip0cm

Fig. 4.  Centrifugal potentials. The abscissa is the distance from the axis of rotation.  The upper curve is the potential associated with a gradient velocity, 
$a^2/r^2$; it appears as a positive contribution to the Hamiltonian and with negative sign in the Lagrangian. The force is the negative of the gradient of the potential,
pointing outwards. 
The lower curve is the negative of the potential associated with solid body rotation. This potential, $b^2r^2/2$, appears with positive sign in the Hamiltonian, and with the same sign in the Lagrangian; the force is in the direction of the gradient, likewise outwards. 
 \b

\ce{\bf The two classical formulations of hydrodynamics}

Fluid mechanics  can be understood as a field theory, and there are two versions,
some times said to be equivalent (Stanyukovich 1960, Sedov 1971), a Eulerian formulation and an alternative `Lagrangian' formulation.  A principal distinction is the following.

In the Eulerian formulation of fluid mechanics  the dynamical variables are the scalar density field and a velocity vector field. This theory becomes a Lagrangian field theory, in the sense of being based on a dynamical action principle,
 only in the case that the velocity field is the gradient of a velocity potential. Thus formulated, this theory cannot describe solid body rotations because the 
 potential $\rho{\vec v}^2/2$  appears with the negative sign in the Lagrangian. Let us be clear about this: the velocity associated with rotational motion in the $x,y$ plane is $\omega(-y,x,0)$ and it is not a gradient, but there is an additional obstacle. If we introduce a term $-\vec v^2/2\, ($or$~ +\vec v^2/2)$ in the Lagrangian we get the right sign in the Hamiltonian but the wrong sign in the equations of motion (or {\it vice versa}).

 The `Lagrangian' version of fluid mechanics handles the centrifugal force correctly, just as particle physics does,  
 but the kinetic energy is in the kinetic part of the Lagrangian. This way one gets the right sign for solid-body motion. This is because the dynamical variables are not $\rho$ and $\vec v$,  but $\rho$ and a vector field usually denoted $\vec x$, 
 related to $\vec v$ by  $ d\vec x/dt=\vec v$. To avoid confusion we shall call it $\vec X$.    This vector field may be interpreted as the position of a particle in the fluid; one chooses an initial value and calls on the equations of motion to predict the future trajectory. But here we are interpreting ${\vec X}$ as a field, with field equations derived from a Lagrangian   with dynamical field variables $\rho$ and $\vec X$. 
  The velocity is the time derivative of the basic field $\vec X$
 and the term $\rho\,\dot{\vec X^2}/2$ is now in the kinetic part of the Lagrangian.
 
  The Navier-Stokes equation is based directly on particle mechanics. But the main equation does not involve the gradient of a Hamiltonian and no Hamiltonian plays a role in the theory. The dynamical field is the velocity but the gradient of no kinetic potential enters the equation.  Instead  the substantive derivative appears,
  $$
  {D\vec v\over Dt} = {\p\vec v \over \p t}+(\vec v\cdot\DD)\vec v.
  $$
  
  Let us look once more at the example of rotary motion. The gradient theory
  describes a motion of the type $\vec v = (a/r^2)(-y,x,0)$, and in this case
  $$
  (\vec v\cdot\DD)\vec v = \DD\vec v^2/2 = \DD(a^2/2r^2).
 $$
 But in the case of solid body  motion, when $\vec v = b(-x,y,0)$ ($a$ and $b$ are constants), we find that
  $$
  (\vec v\cdot \DD) v = -\DD\vec v^2/2= \DD(-b^2r^2).
 $$
  
The Navier-Stokes equation is based on particle mechanics and both kinds of motion are treated correctly, but there is no Hamiltonian; this is because the velocity is the  primary dynamical variable. To get beyond this stage, to set up a Lagrangian theory of non potential flow, it is necessary to introduce the field $\vec X$, as in
the `Lagrangian' version of fluid mechanics. Traditional `Lagrangian hydrodynamics' can be formulated as an action principle but  since it does not
incorporate the equation of continuity, it is not hydrodynamics; a scalar velocity potential is needed as well.

When both flows are present, as in
$$
\vec v = ({a\over r^2} +b)(-y,x,0)
$$
we find that there is a cross term,
$$
  (\vec v\cdot \DD) \vec v = \DD\bigg( 
{a^2\over 2r^2} - {b^2r^2\over 2} + ab\ln r^2 \bigg)\eqno(4.4)
$$

\bb

\no{\bf V. Action principle for general flow}
 
 With the Lagrangian density that we have used for potential flow as a starting point,
 we begin by adding a new term,
 $$
\L = \rho\Big(\dot\Phi+ {1\over 2}\dot{\vec X}^2 -{1\over 2}(\DD\Phi)^2  
\Big)-f-sT.\eqno(5.1)
$$
The term $\dot{\vec X}^2/2$  appears with the sign that is appropriate for solid-body rotational motion. It is a contribution to the kinetic part of the Lagrangian density.

We digress  to show that this Lagrangian has precedents. The problem is to understand the flows that are observed in Couette flow of a superfluid.   The potential term \break
$ -(\rho/2)(\DD\Phi)^2 $ may be considered as a part of the free energy density; the new term $(\rho/2)\dot{\vec X}^2$ plays exactly the same role as the second term in
$$
F' = F-M\Omega.\eqno(5.2)
$$
where $M = I\Omega,\, I$ the total moment of inertia. This equation appears in the work of Hall and Vinen (1956) (and in Landau and Lifshitz 1955, 1958); $F$ is the free energy, including the kinetic energy due to potential flow, $M$ is the total angular momentum and $\Omega$ is the angular velocity.  This quantity $F'$ is required to be a minimum with respect to variation of $\Omega$, with $I$ fixed, and the resulting equation is the same as when $I\Omega^2/2$ is varied with $M$ held fixed. The context is superfluid Helium but the theory
applies just as well to normal fluids. The physical interpretation as well as the mathematical structure is thus precisely the same as that of the above tentative Lagrangian (5.1). This explains the negative sign: $F'$ is not a  modified free energy but a Lagrangian! End of digression.

The Hamiltonian density is now
$$
{\cal H} = {\rho\over 2}\dot {\vec X}^2  + {\rho\over 2}(\DD\Phi)^2+f+sT. \eqno(5.3)
$$
This is correct for all flows, most notably for solid body rotation,
$\dot{\vec X} = b(-y,x,0)$. 

We shall compare this theory with the Navier-Stokes equation, but we first need to add another term to the Lagrangian, suggested by the $ab$ term in Eq. (4.4), the final form being
$$
\L = \rho\Big(\dot\Phi + {1\over 2}\dot {\vec X}^{\,2} +\kappa\dot{\vec X}\cdot\DD\Phi- {1\over 2}(\DD\Phi)^2
\Big) -f - sT.\eqno(5.4) 
$$
Variation of $\Phi$ will give the equation of continuity. The factor  $\kappa$ is an adjustable, dimensionless  parameter.
The expression (5.3) for the Hamiltonian is not affected.
\b

Let us list all the Euler - Lagrange equations.

1. Variation of $T$: the adiabatic relation, as always.

2. Variation of $\Phi$: the continuity equation, now 
$$
\dot\rho+\DD\cdot \Big(\rho\kappa \dot{\vec X}- \rho\DD\Phi\Big) = 0.\eqno(5.5) 
$$
The unique velocity of mass transport is therefore  
$$
\vec v:= \kappa\dot{\vec X}-\DD\Phi.\eqno(5.6)
$$
This is what corresponds to the velocity that appears in the Navier-Stokes equation.

3. Variation of the vector field:
$$
{\p\over \p t}\vec m = 0,~~~\vec m :=
 \rho (\dot{\vec X} + \kappa \DD\Phi) = 0.\eqno(5.7) 
$$
 This will be referred to as the momentum equation, so called because of an alternative interpretation of the field $\vec X$ as the position coordinate of a particle in the fluid,
 or as the coordinate of a point on a vortex line. The ``momentum", in this interpretation,
 is the variable conjugate to the ``translation" - interpreted as a translation of a vortex line -   (Lund and Regge 1976)  $\vec X\rightarrow \vec X + \vec a$.
 Our approach extends this interpretation to the case of of a continuous distribution of vorticity. Incidentally,  Lund and Regge clearly 
disassociate the `momentum' from the mass flow. 
   
 If the absence of a source term for the vector field $\vec m$ seems unusual, it is of interest to note that, in the cited studies of Couette flow in superfluid Helium, the solid-body flow was usually approximated by a constant, not as an independent dynamical variable. 
In the alternative interpretation, in terms of vortex lines, the momentum is constant  because the vortex line moves with the liquid. See below, Eq. (6.1).
A source for the field $\vec X$ will be introduced later.

4. Variation of the density:
$$
\dot\Phi + {1\over 2}\dot {\vec X}^{\,2} + \kappa \dot{\vec X}\cdot\DD\Phi 
-{1\over 2}(\DD\Phi)^2 -{\p f\over \p \rho}-ST = 0.\eqno(5.8)
$$
These variational equations are conservation laws; together they imply the conservation of energy, ${\cal H}$ in (5.3) being the energy density.
 Note that two independent vector fields, $\dot{\vec X}$ and $-\DD\Phi$, are needed to get the correct signs 
for both squared velocity terms; opposite signs in (5.8), both positive in the Hamiltonian
density. The term
$\kappa\dot{\vec X}\cdot\DD\Phi$ does not contribute to the Hamiltonian, but it is crucial
for the equation of continuity. 

The stationary flow that is the main object of this study is a solution of the Euler-Lagrange equations. Numbers 2 and 3 are solved by  taking $\rho$ to depend on $r$ only, $\dot \Phi$ constant, and 
$$
\Phi = t\dot\Phi +a\theta,~~~ \DD \Phi = {a\over r^2}(-y,x,0),~~~
 \vec X= bt(-y,x,0).\eqno(5.9)
 $$
When these solutions are inserted into (5.8) we get, in the case of an ideal gas,
for stationary flow,
 $$
{a^2\over 2r^2} - {b^2\over 2}r^2 + \kappa ab = C-(n+1){\R} T.\eqno(5.10)
$$
This is the equation that we can compare with the Navier-Stokes equation and it is very nearly the same, but to compare we must take the gradient.
$$
\rho\DD\bigg({a^2\over 2r^2} - {b^2\over 2}r^2 +\kappa ab\bigg) = -\DD p.
$$
With the same flow, the Navier-Stokes equation is
$$
\rho\DD\bigg({a^2\over 2r^2} - {b^2\over 2}r^2 +  ab\ln r^2\bigg) = -\DD p-\nu\DD\Delta\vec v.
$$
Of course, we cannot reproduce the viscosity term in the Euler-Lagrange equations
(but see below). The cross term $ab$ is a constant, while in the Navier-Stokes equation it is replaced by the logarithm, Eq(4.4).

\vskip1cm

With the lagrangian density (5.1) a significant step in our program has been completed.  We now know that the idea of a viable Lagrangian formulation of thermodynamics is not absolutely limited to potential flows. This leaves many things yet to be done, for example, the complete interpretation of the two vector fields that make up the total flow,  the integration with electromagnetism, and with General Relativity. First of all we need to apply the new, tentative insight to a range of phenomena; all that we know at present is that our new Lagrangian density is suitable for cylindrical Couette flow in the laminar regime. A closely related phenomenon, the flow along plane, moving walls (linear Couette flow) will be examined in Section VII. Note that the new theory is  new only in a context where both velocities are present; when one or the other is absent there is perfect agreement with the traditional approach. 

The appearance  of an extra variable, in the action principle for general flows, should not be surprising. The work of Feynman (1954), and of Landau and Lifshitz (1955,
1958) on the application of statistical mechanics to liquid Helium is based on a separate and 
different treatment of two types of excitations, the `phonons' being accounted for by a 
gradient vector field and the `rotons' understood as a different degree of freedom.
These papers, and especially the much quoted paper by Hall and Vinen (1956) that they inspired, briefly refer to an action principle. What is minimized is a function  $ F' = F-I\Omega^2/2$, where the second term can be identified with  the term 
$\rho\dot{\vec X}^2/2$ in our Lagrangian. (The term $-\rho(\DD\Phi)^2/2$  of our
Lagrangian density is included in the free energy $F$.)  In most of the papers that followed, the strong hint of a Lagrangian structure was systematically suppressed.
See however, the more recent review by Fetter (2009). The treatment of vortex lines referred to above also makes use of two different velocity fields although in this case it is the field $\dot{\vec X}$ that occupies center stage.

The field $\dot{\vec X}$ is needed because the flow velocity  is not always irrotational.
 In the special case of certain vortices the curl of the velocity field is zero
everywhere except on the vortex line. In that case it would be possible to formulate
a theory (choose a gauge) where ${\vec X}$ is concentrated on the vortex line. This idea was 
developed by Regge and others, in papers already quoted in the introduction.
\bb

\no{\bf VI. The viscosity}

Most of the equations of motion represent conservation laws. In this respect we are in full agreement with the dominant approach to non equilibrium thermodynamics. 
 Eq.(5.5) is the condition that the total mass be conserved.   Eq. (5.6) can be interpreted as the conservation of  `momentum'. The Navier-Stokes equation relates the non conservation of momentum to the viscosity. To apply the theory to the case that the viscosity is not zero we  add a source term to Eq.(5.7), 
$$
{\p \over \p t}\Big(\rho \dot{\vec X} +\kappa\rho\DD\Phi\Big) = \mu\Delta \vec v,~~~
\vec v := \kappa\dot{\vec X}-\DD \Phi.\eqno(6.1) 
$$
To compare this with the Navier-Stokes equation we must combine it with the gradient of Eq. (5.8) (multiplied by $\rho$) to convert it to an equation for
the time derivative of the field $\vec v$,  
this results in very close term-by-term agreement,  including the viscosity term.  

The addition of an extra term to the momentum equation is legitimate; it represents the
influence of another degree of freedom that, because it is characterized by a longer time scale, can be approximated as a quasi-static development. The  dissipated energy is transferred to the other degrees of freedom.

When the system of Euler-Lagrange equations is modified by the addition of the viscosity term in (6.1) the energy is no longer conserved, instead
$$
{d\over dt}{\cal H} = \dot{\vec X}\cdot \mu \Delta\vec  v.\eqno(6.2)
$$
This  is not a complete description of the dissipative process; Eq.(6.2) was obtained under the condition that the entropy remains fixed, which is possible but not known to be true.

\bb

\no{\bf VII. Comparing two approaches}

 \ce{\bf  Predicting the flows. }
A major success of the application of the Navier-Stokes equation to laminar, cylindrical Couette flow is the prediction, on the basis of the tangential component
(4.2): $\mu \Delta \vec v = 0$, of just two kinds of flow, in this case potential flow and solid-body flow. Intuitively, the latter is a little unexpected; this flow is attained, or nearly attained, when the outer cylinder is driving and the inner cylinder is slipping. The Euler Lagrange equations associated with the Lagrangian density (5.4)  place no restrictions on the factor $b$ in 
the solution $\vec X=b(-y,x,0)$, but the inclusion of  the viscosity as in Eq.(6.1)
effectively implies that $\Delta \vec v = 0$ for stationary flows, reducing $b$ to a constant. The approach to viscosity is thus precisely the same in both theories. A feeling that the underdetermination of the system of equations in the absence of viscosity, present in both approaches, is something to worry about, persists. It is reinforced by the fact that the Lagrangian density does not contain any space derivatives of the field $\vec X$. This problem cannot be resolved in the non relativistic context. See the last section of the paper.
 \bb

\ce{\bf   The density profile. } The Euler Lagrange equation obtained by variation of the density, Eq.(5.8), reduces in the case  that the flows are as in (5.9) to Eq. (5.10).
This is in agreement with the Navier-Stokes equation except that the cross term $ab$
is constant, while in the Navier-Stokes equation it depends logarithmically on the radius.
If the angular velocity $b$ were to vary with $r$, the term
$\vec X\cdot \DD \Phi$ in the lagrangian density would no longer be  a constant.  If $b$
were replaced by $br^{-\tau}$ we would obtain
 $$
{1\over 2}\dot {\vec X}^{\,2} + \kappa \dot{\vec X}\cdot\DD\Phi 
-{1\over 2}(\DD\Phi)^2={a^2\over 2r^2} - {b^2\over 2}r^{2(1-\tau)} +ab\kappa r^{-\tau} = C-(n+1){\R} T.\eqno(6.1)
$$
See Fig. 5. This is in substantial agreement with  Navier-Stokes in the limit of small $\tau$. And it offers an extra free parameter for a small amount of fudging.  In the case that the flow is either pure potential or pure solid-body there is full agreement.

\epsfxsize.5\hsize
\centerline{\epsfbox{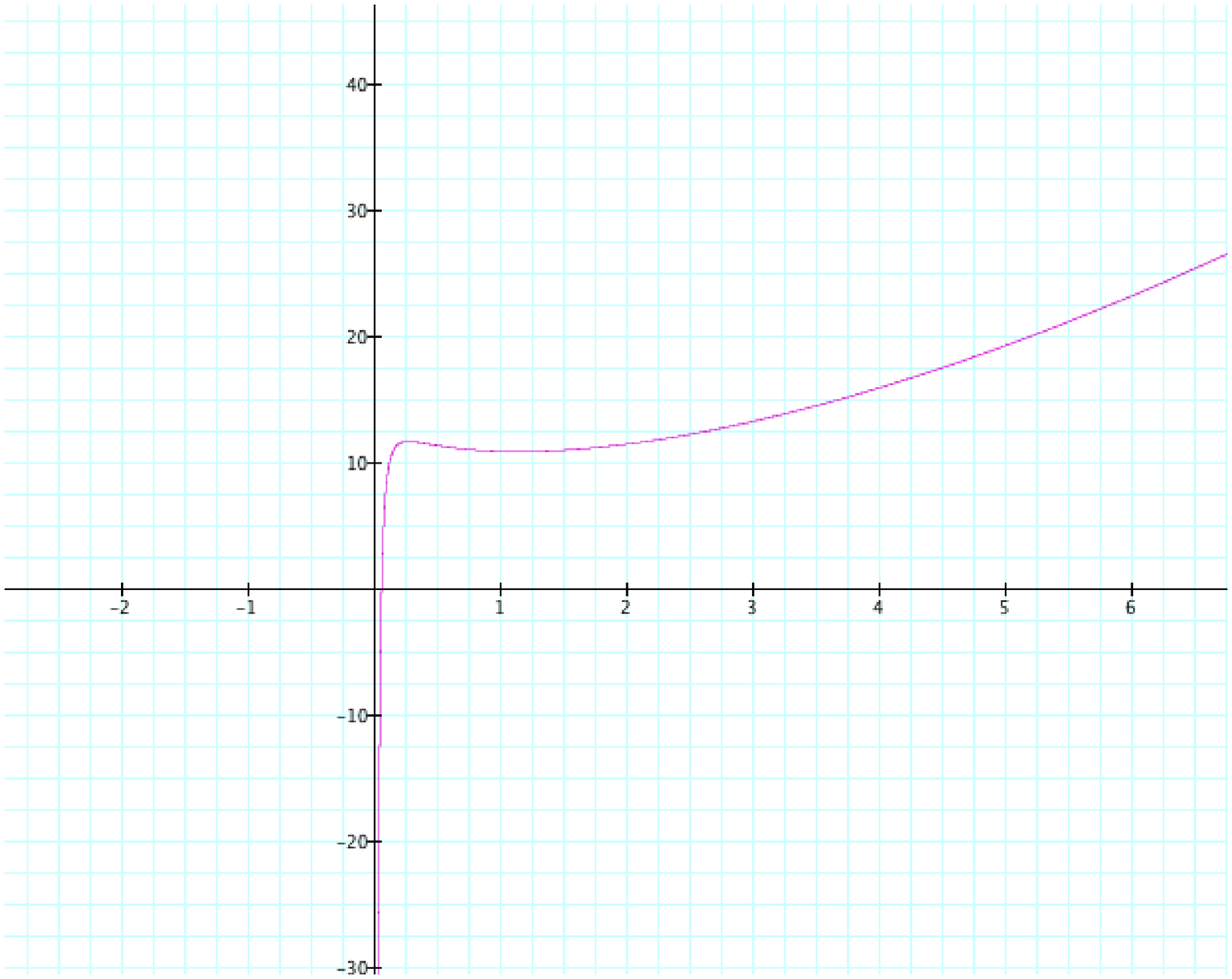}}
\vskip0cm

 Fig. 5. A temperature profile for cylindrical Couette flow. The small dip for small
radius  is produced by the $ab$ term in  (6.1).
A similar feature has been observed. 
 
\b

\ce{\bf   The energy.}
A principal reason for bringing out an alternative to the approach that relies
heavily  on the Navier-Stokes equation is of course the improved position of the energy concept.
The equations of motion do not differ greatly from the equations used in the traditional  method but, instead of an {\it ad hoc} formula for ``energy" that is required to be conserved as an additional postulate (See for example M\"uller 2007), we have a first integral of the equations of motion. We note that some authors require that the ``energy equation" hold in
consequence of the other conservation equations, thus expressing a point of view similar to ours. (Khalatnikov 1956.)

\b
\ce{\bf Instabilties}

The main interest in Couette flow, from the very beginning, has been the question of
instabilities of the laminar flow that is observed when the speed of rotation is increased. One approach to stability is perturbation theory. The unperturbed motion is of the form of `solid-body' rotation, with fixed angular velocity  (Fetter 2009) or a combination of this with potential flow (Andereck et al 1986). Usually, the  perturbation is assumed to affect only the potential field;  this field being the only dynamical variable.
This system is characterized by a Fetter-Walecka Lagrangian  and it was treated as such in the review by Fetter already quoted. From our present perspective,  a complete treatment should  include a variation of both vector fields. The success of the application of the Navier-Stokes equation to the stability of flow in an incompressible fluids does not extend to thermodynamical systems defined by an equation of state. This may be related 
to the fact that the interaction between kinetics and heat is carried by the potential 
component of the flow. We propose that a more general approach to stability problems
may give some new insight.

Let us emphasize that two velocity fields already appear in traditional treatments of Couette flow, though one of them is frozen at the fixed, unperturbed value or else not mentioned at all. this may be a valid approximation in certain cases. Our suggestion is that, in general, it is advantageous to treat both fields as independent field variables.

 Here is a very preliminary attempt to extend our model to an investigation of   instabilities.The expression $(n+1){\R} T$ on the right hand side of (5.10) is not valid at very low temperatures,  but in the case of an ideal gas, with uniform entropy, 
it is proportional to a power of the density and it is perhaps reasonable 
to expect it to go to zero with the density. In any case it will take a fixed, numerical value when the density becomes zero. That would suggest local cavitation,
a likely trigger of instabilities.  Thus the onset of local turbulence may be
expected  to occur for a fixed value of this quantity.
$$
\dot{\vec X}^2/2+\kappa \dot{\vec X}\cdot\DD\Phi - \DD\Phi^2/2 = 
{1\over 2(\kappa^2+1)}\big(\vec u^2-\vec v^2+2\kappa \vec u\cdot \vec  v\big) = {\rm constant}.\eqno(6.2)
$$
where $\vec u = \vec m/\rho =  \dot{\vec X}+\kappa \DD\Phi$ and $\vec v = \kappa\dot{\vec X} -\DD\Phi$.

In the $a,b$ plane this resembles a hyperbola open to the positive $b$ axis. 
This too is in qualitative accord with experiments. See Fig.s 6 and 7.
\b

\ce{\bf Brenner's bi-velocity theory}

This theory was invented to account for the fact that several different velocity fields are needed to describe certain fluids, especially fluids that carry electric charge.
The flow of charge is not necessarily the same as the flow of mass (Brenner 2011,
Brennerrenardy 2013). In that case the recognition of two densities leads
naturally to two flow velocities. A concept of dual velocities is a feature of General Relativity as well,  as in Weinberg's treatment of conservation laws (Weinberg 1972). The theory developed in this paper also requires two different kinds of velocity fields, 
but the interpretation is different and as yet not completely understood. Perhaps most important is the fact that the number of degrees of freedom, including the density,
is four, as in classical hydrodynamics.
We have seen that the velocity associated with momentum flow is different from 
the velocity of mass flow, but that observation is somewhat formal. For other references to bivector velocity fields see the quoted review by Martin (2010).

 \ve
   
\epsfxsize.5\hsize
\centerline{\epsfbox{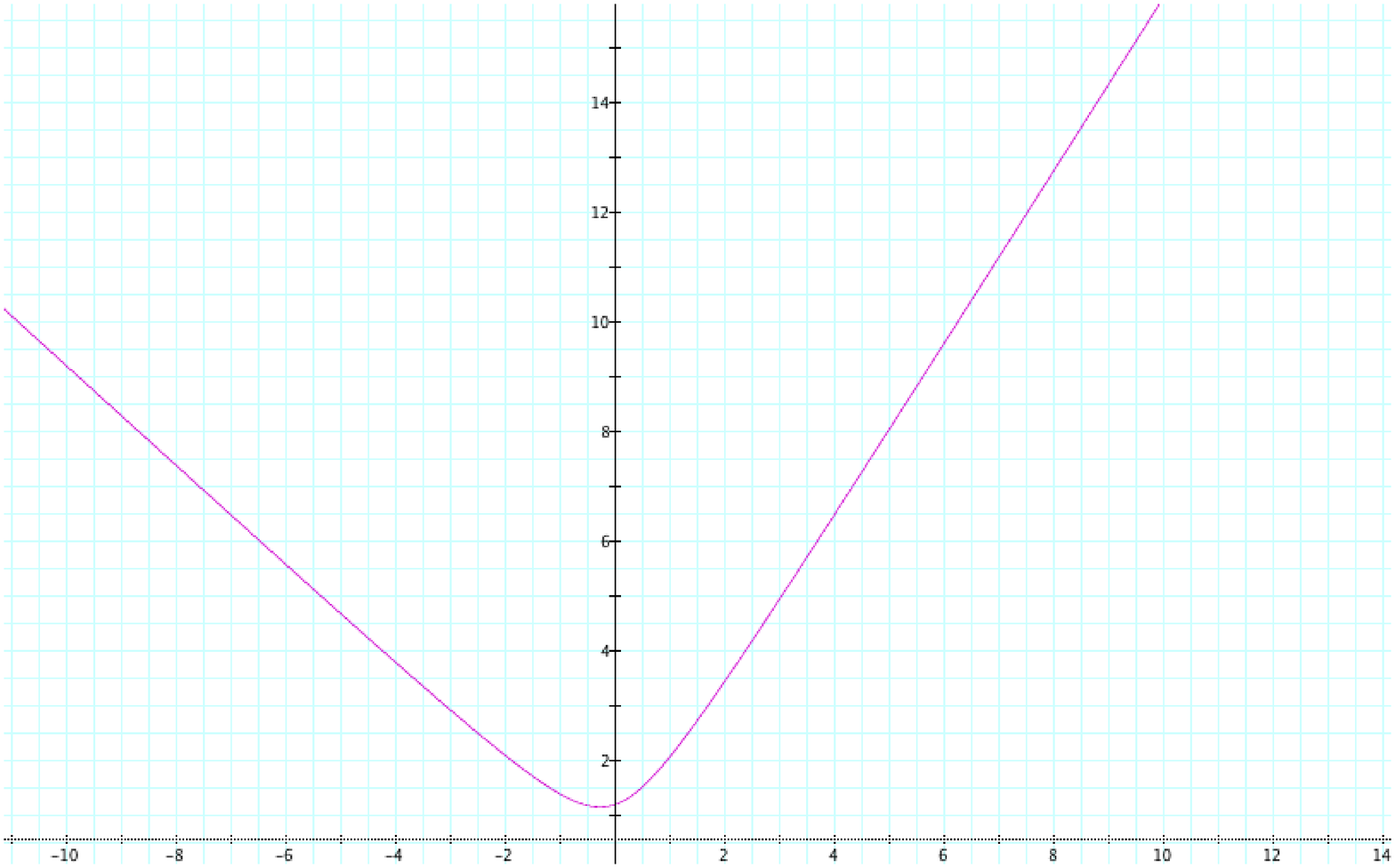}}
\vskip0cm

Fig.6. This is the locus, in the plane with coordinates $\omega_o/\omega_i$ = angular velocity of outer/resp. inner cylinder, of equation (6.2) for a fixed value of $C_0$.
 The parameters are the same as in Fig.7 and the best value for $\kappa$ is -1/3.

\bb \bb

\epsfxsize.7\hsize
\centerline{\epsfbox{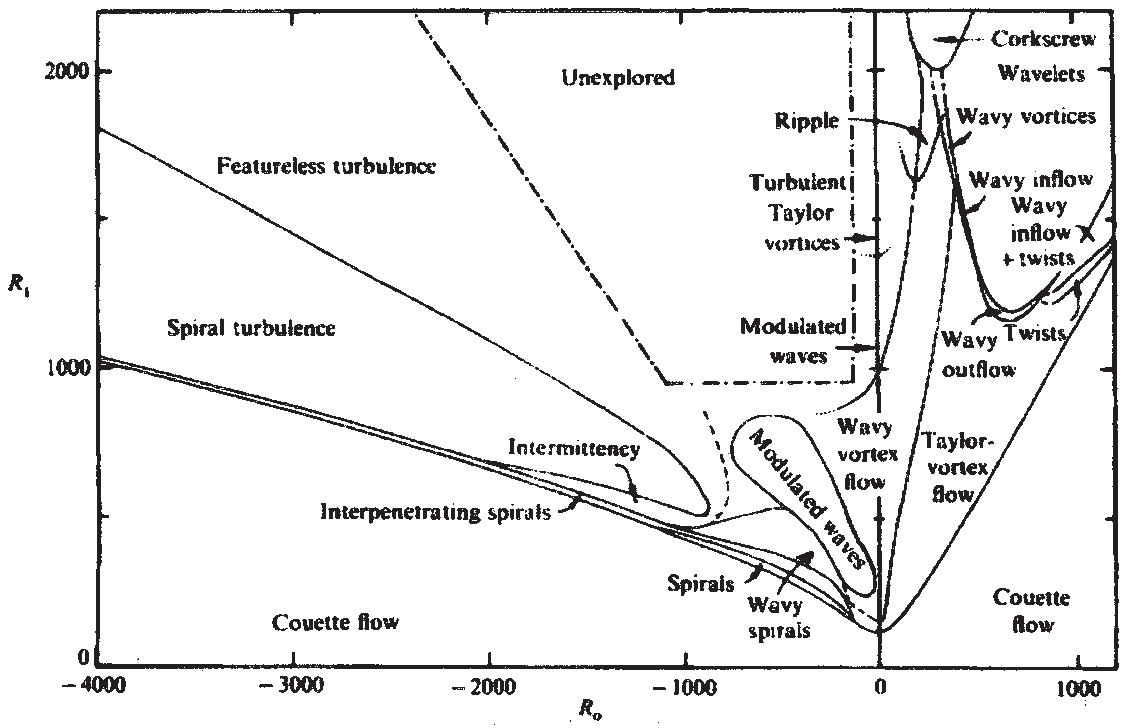}}\
\vskip0cm

\vskip0in

Fig.7. The experimental envelope of stability as per Andereck et al. The ratio between inner and outer cylinder radii is .83.

\bb

\no{\bf VIII. Couette flows between plane walls} 
 
 This example will  show that one cannot expect the new action principle to agree with the Navier-Stokes equation in all cases, even in the limit of negligible viscosity.
It is an example where the Navier-Stokes equation has very
limited predictive power beyond the identification of the two principal modes of flow.  
 
 Here again is the Navier-Stokes equation,
 $$
\rho\Big(\dot{\vec v} + (\vec v\cdot \DD) \vec v\Big)= - \DD p - \mu \Delta  \vec v.\eqno(8.1)
$$
Consider the problem of flow in the space 
bounded by plane walls parallel to the $x,z$-plane, $y= y_0$ and $y= y_1$.
We limit our attention to stationary flows parallel to the $x$ axis,
$$
\vec v = (v,0,0),
$$
with the function $v$ depending only on $y$. 

What the Navier-Stokes equation has to say is this. Projected on the direction of the flow, the velocity and the pressure are homogeneous (constant), and the equation  reduces to 
$$
\mu\Delta v_x = 0.
$$
with the general solution, when $\mu \neq 0$, 
$$
\vec v = (a +by)(1,0,0),~~~a,b ~~~{\rm constant}.\eqno(8.2)
$$

We are studying stationary flows, with $\dot{\vec v}=0$; and  since the velocity is constant along the direction of motion, 
$$
{D \vec v\over D t}:=\dot{\vec v}+(\vec v\cdot \DD) \vec v = 0.\eqno(8.3)
$$
Consequently, all that remains of the Navier-Stokes equation is this:
$$
0= - \DD p - \mu \Delta  \vec v,\eqno(8.4)
$$
projected on the $y$-axis;  the pressure gradient is balanced against the viscosity.
Although the kinetic energy - if it is relevant to invoke energy in the context of the Navier-Stokes equation - varies with the position, there is no force associated with this variation of the kinetic energy. In the special case that the flow is potential the kinetic energy is uniform,  
 but the absence of a kinetic force when $b\neq 0$ tells us that this other type of flow is of a very different character, since 
a non zero gradient of the kinetic energy in any flow was expected to generate  a force. Perhaps this is behind some hints found in the literature, to the effect that, in the context of the Navier-Stokes equation, the relation between linear flow and cylindrical flow is not perfectly understood  (Faisst and Eckhardt 2000).

What is the meaning of (8.4)? One may, for example, assume that the pressure is that of an ideal gas,
$$
p={\R}\rho T,
$$
but since the temperature is very rarely measured,  perhaps the polytropic relation will be used instead,
$$
p \propto \rho^\gamma.
$$
Unfortunately the pressure also is rarely measured with the required precision.  Consequently, it is very rare that an analysis is made in the interest of verifying a theory and all that can be said is that experimental results are not known to contradict the Navier-Stokes equation. Instead, they serve to measure the viscosities.
	
It is unfortunate that few experiments on Couette flow are made with the idea of studying the laminar flow;  the aim is usually to determine the conditions under which it breaks down to be replaced by more complicated flows including turbulence. Our purpose, so far, has been more modest, since we aim to understand laminar flow only, 
but more ambitious since we are looking for an application of a general theory with definite precepts including as far as possible a Lagrangian variational principle. The most urgent question is always the simplest: can we understand the observed laminar flow in the limiting case in which the role of viscosity is negligible?

\epsfxsize.5\hsize
\centerline{\epsfbox{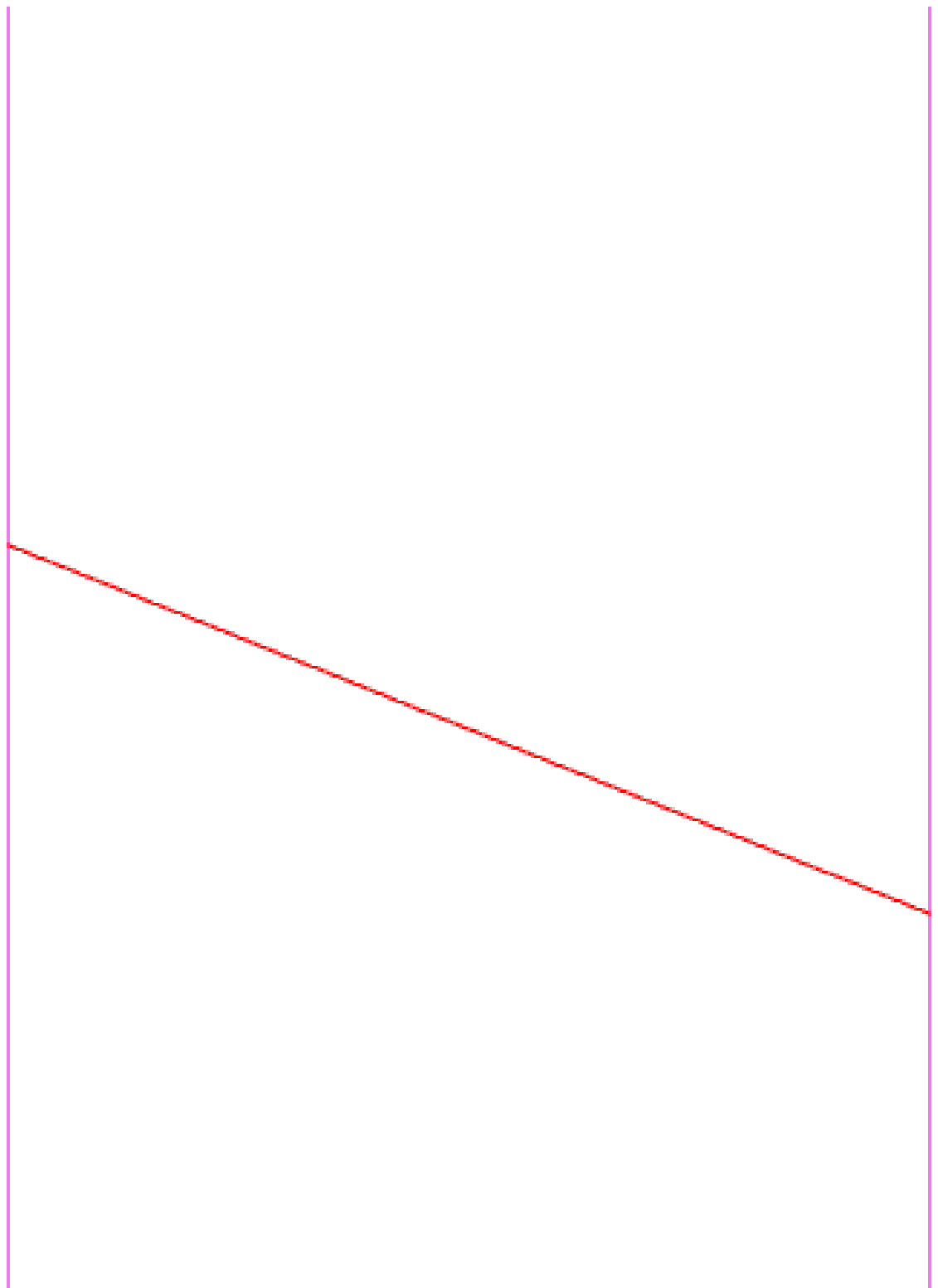}}
\vskip0cm

Fig. 8.  A velocity profile of linear Couette flow according to the variational principle.   
\b
 
For simplicity, from here onwards, in this section,  we set the parameter $\kappa $ at unity. Our theory, in the present state of development, includes an equation of continuity
that, when the flow has the form (8.2), illustrated in Fig. 8, namely
$$
\vec v =  \dot{\vec X}-\DD\Phi = (a +by)(1,0,0),\eqno(8.5)
$$
demands that the density depend on the coordinate $y$ only. Solutions for the vector and scalar fields include the following
$$
-\DD \Phi = a(1,0,0),~~~\vec X = bt(y,0,0).\eqno(8.6)
$$
The field equation  is satisfied,
$$
\p_t(\rho \dot{\vec X} +\DD \Phi) = 0,
$$
and finally there is  the equation that comes from variation of the density; in the case of an ideal gas,
$$
{a^2\over 2} - aby-{b^2y^2\over 2}= C-(n+1){\R} T\eqno(8.7)
$$
or
$$
{1\over 2}(a+by)^2 = a^2  -C+(n+1){\R} T.
$$

 This contrasts with the Navier-Stokes equation (8.4).  The fluid is contained between moving walls located at two values of the coordinate $y$. If this interval includes the  
plane with coordinate $y=-a/b$, then the temperature has a minimum there, as in Fig. 9.

\epsfxsize.5\hsize
\centerline{\epsfbox{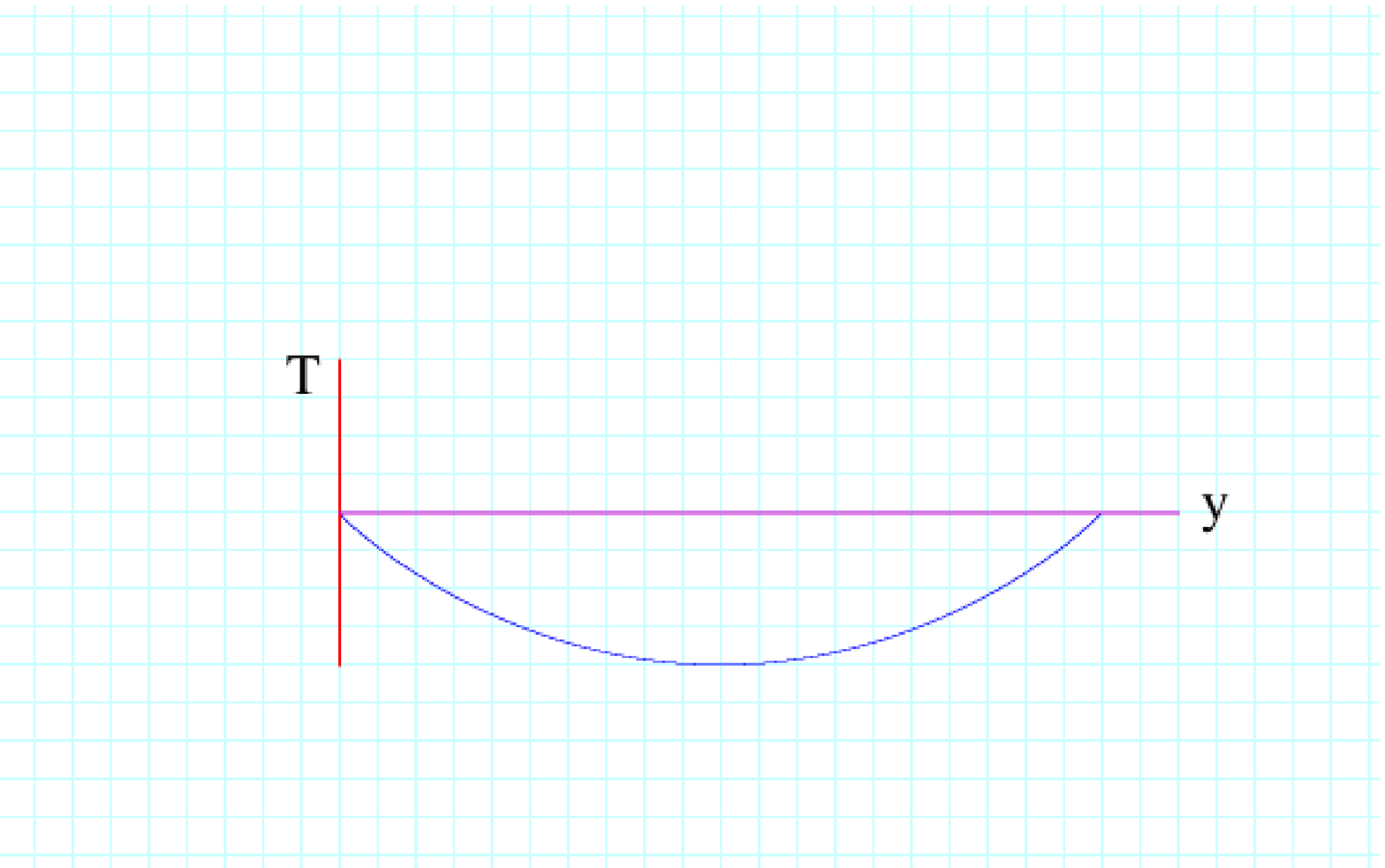}}
\vskip0cm

Fig. 9. The temperature profile of linear Couette flow between two walls
\b
 
 The order of magnitude of the variation of temperature is that of $v^2/\R$. If
the velocity is  1 m/sec this will imply a temperature variation of order 
$10^4/C_p$, a very small variation indeed. So it is unlikely that this temperature variation has been observed.

It is nevertheless significant that, in remarkable contrast with the Navier-Stokes approach, our approach leads to a definite, quantitative prediction without the explicit intervention of viscosity, valid because the role of viscosity is merely to 
restrict the allowed velocity  to the form (8.6). 

Stationary flow past a flat or nearly flat surface continues to be a vital subject
in ship design; see the recently evolving subject of stepped hulls. 
 
\bb

\no{\bf IX. Final remarks}

\no{$\bullet$} This paper was motivated by the need for an action principle formulation of General Relativity in the presence of matter, which implies an action principle for thermodynamics. An action principle for 
thermodynamics has  been presented that, to encompass general velocity fields, relies on the introduction of two kinds of velocities. In hindsight there were many indications
that this dual approach to fluid velocity was needed.

\no{$\bullet$} The decision to make use of the Lagrangian velocity $\dot {\vec X}$ in 
hydrodynamics has startling consequences for relativistic physics. 
A precept that has dominated all relativistic fluid dynamics is that the
relativistic 3-vector velocity field  shall be upgraded to a relativistic 4-vector 
velocity field. (Remember that the electric field is a 3-vector but is promoted to a tensor field.)
Attempts to imbed the field $\dot{\vec X}$ in a tensor field has led  to the following field structure.

Consider a relativistic 2-form $Y$ and set
$$
{\vec X}^i = {1\over 2}\epsilon^{ijk}Y_{jk},~~~\eta_i = Y_{i0},
$$
with latin indices running over 1,2,3. Then the Lorentz invariant scalar
$$
dY^2 = {1\over 2}\bigg(\dot{\vec X} + \DD\w \vec \eta\bigg)^2 - 
{c^2\over 2}(\DD\cdot \vec X)^2
$$
is invariant under the gauge transformation $\delta Y=d\xi$. There is a family of gauges  in which the field $\vec \eta=0$ and this expression reduces to $\dot{\vec X}^2/2$.
Other terms in the Lagrangian also can be made manifestly Lorentz invariant,
while $\dot\Phi - \DD\Phi^2/2$ is the non relativistic limit of 
$(1/2)(g^{\mu\nu}\psi_{,\mu}\psi_{,\nu}-c^2),$ with $\psi = c^2t+\Phi$ (Fronsdal 2007). In this form the theory becomes a relativistic Lagrangian field theory and the associated energy-momentum tensor is
an appropriate source for Einstein's equation in the presence of a rotating source,
a source that satisfies the Bianchi constraint (Fronsdal 2015). 

\no{$\bullet$} An unexpected effect of moving up to the relativistic context is an additional field equation related to the variation of the action with respect to the gauge field $\eta$. The gauge is fixed in the non relativistic theory but variation of the gauge field nevertheless furnishes an additional Euler-Lagrange equation,
$$
\DD\w (\rho\dot{\vec X}+\rho\kappa\DD\Phi) = 0.
$$
In the 3-dimensional context this is not a field equation but a constraint on a gauge theory that is difficult to understand within the gauge-fixed, non-relativistic context. We shall not enter this topic here but refer to another paper
(Fronsdal 2015)

\no{$\bullet$} A relativistic electromagnetic theory of fluids was created by Minkowski (1908) using the 3-vector-to-4-vector prescription. If we are going to use the 
velocity field $\dot{\vec X}$, then the theory of relativistic electromagnetism
of fluids will have to be recreated to accommodate.  We suggest that an alternative treatment of rotators in a magnetic field  is a way to begin. Or else: What is the effect of rotation on the operation of a car battery?

\no{$\bullet$} Finally it is clear that the proposed development of the structure of velocity space may have a profound effect on hydrodynamic stability studies, especially 
where it is needed most: in the case of compressible fluids where the subject is thermodynamics rather than just hydrodynamics. 

$\bullet$ The main suggestion advanced in this paper is that two types of velocity fields
have to be treated as independent dynamical variables. This was not normally done, but we 
alluded to some important exceptions. In a work inspired by Onsager 
(Eyink and Sreenavasen 2006), Rasetti and Regge (1975) and Lund and Regge (1976) have used the field $\dot{\vec X}$ in their description of the internal and external dynamics of a vortex string. The curl of this field is zero except along the vortex line.  The connection to the Nambu-Goto string is further developed by Zheltukhin (2014).  Here we find the justification for referring to the field $\vec m$ - 
Eq.(5.7) - as the ``momentum". Momenta are conjugate to translations; in this case the translation of the vortex line. In these papers 
a Lorentz invariant extension is also developed; it involves the same 2-form that is mentioned above.

$\bullet$. Finally, retirening to the starting point, here is the principal part of relativistic action 
that provides Einstein's equation with a source of rotating matter,
$$
\int d^4x\sqrt{-g}\big(R + \L\big),
~~~~\L = {\rho\over 2}g^{\mu\mu'}g^{\nu\nu'}g^{\lambda\lambda'}dY_{\mu\nu\lambda}dY_{\mu'\nu'\lambda'}.
 $$
 The energy momentum tensor is very different from that advocated by Tolman,  the main advance is that the Bianchi identity as well as the equation of continuity are satisfied.

 \bb
\ce{\bf Acknowledgements}

I thank Tore Haug-Warberg at NTNU in Trondheim and many colleagues at the Bogoliubov Institute of JNIR at Dubna for stimulating discussions. This work was initiated at NTNU
in 2013 and continued at JINR in 2014. The hospitality of both institutions and their
staffs is gratefully acknowledged. The warm hospitality of Vladimir Kadyshevsky 
of the Bogoliubov Institute and the support of JINR are greatly appreciated.  I thank
Alan Weinstein, Elliott Lieb and Mark Raizen for hospitality.
\vskip1cm

\bb 
 
\ce{\bf REFERENCES}

\noindent Andereck, C.D.; Liu, S.S. and Swinney, H.L.``Flow regimes in a circular Couette 

system with independently rotating cylinders". J.Fluid Mech. {\bf 164} 
155-183  (1986).   


\noindent Brenner, H., 
``Proposal of a critical test of the Navier-Stokes-Fourier paradigm for 

compressible fluid continua", Phys.Rev. E 87, 013014 (2013).

\noindent  Brenner, H., ``Beyond the no-slip boundary condition",

 Physical Review E 84, 046309 (2011). According to this paper, N-S works only for 
 
 incompressible fluids.
 

\noindent Couette, M., ``Oscillations tournantes d'un solide
de révolution en contact avec un fluide

visqueux," Compt. Rend. Acad. Sci. Paris
{\bf 105}, 1064-1067 (1887).

\noindent Couette, M., ``Sur un nouvel appareil pour
l'\'etude du frottement des fluides", 

Compt.Rend. Acad. Sci. Paris {\bf 107}, 388-390 (1888).

\noindent Couette M., ``La viscosit\'e des liquides," 

Bulletin
des Sciences Physiques {\bf 4}, 40-62, 123-133, 262-278 (1888).  

\noindent Couette, M., ``Distinction de deux r\'egimes dans
le mouvement des fluides," 

Journal de
Physique [Ser. 2] IX, 414-424 (1890)

\noindent Couette, M., ``Etudes sur le frottement des
liquides," 

Ann. de Chim. et Phys. [Ser. 6] 21, 433?510 (1890).

\noindent  de Socio, L.M., Ianiro, N. and  Marino,L., ``Effects of the Centrifugal Forces on a Gas

Between Rotating Cylinders", J. Thermophysics and Heat Transfer, {\bf 14}    (2000). 

\noindent Dukowicz,J.K., Stephen F.,  Price, S.F. and  Lipscomb,  W.H.,
``Consistent 

approximations and boundary conditions for ice-sheet
dynamics from a principle of 

least action",
Climate, Ocean and Sea-Ice Modeling (COSIM) Project, Group T-3, 

MS B216, 
Los Alamos National Laboratory, Los Alamos,
New Mexico 87545, USA. 

Journal of Glaciology - J GLACIOLOGY , vol. 56, pp. 480-496 (2010).
 
 
\noindent Emden, R.,  {\it Gaskugeln}, Teubner, Wiesbaden, Hesse, Germany, 1907.

\noindent Eyink,  G.L. and Sreenivasan, ``Lars Onsager and the theory of hydrodynamic turbulence 

Rev. Mod. Phys., Volume 78, (January 2006). 

\noindent Faisst, H. and Eckhardt, B., ``Transition from the Couette-Taylor system to the 

plane Couette system",
 arXiv:physics/0003102v1 [physics.flu-dyn] (30 Mar 2000).

\noindent Fetter, A.L. and Walecka, J.D., {\it Theoretical Mechanics of Particles
and Continua}, 

MacGraw-Hill NY (1980).

\noindent Fetter, A.L., ``Rotating trapped Bose-Einstein condensates", 

Rev.Mod.Phys. {\bf 81} 647-691  (2009).

\noindent Feynman, R.P.,  
``Atomic Theory of the Two-Fluid Model of Liquid Helium",

Phys. Rev. {\bf 94}, 262-277 (1954).

\noindent Feynman, R.P., in {\it Progress in low temperature physics 1}, Chapter II, page 36, 

C.J. Gorter Ed., North Holland (1955).

\noindent Fronsdal, C., ``Ideal Stars and General Relativity", Gen.Rel.Grav. 39 1971-2000 (2007).

\noindent Fronsdal, C., {\it Thermodynamics of fluids, a monograph}, in progress (2014a).

Available at fronsdal.physics.ucla.edu.

 \noindent Fronsdal, C., ``Heat and Gravitation.  The Action Principle",  
  Entropy  {\bf 16}, 1515-1546 
  
  (2014b).
doi:10.3390/c16031515. (arXiv 0812.4990v3, revised)

\noindent Fronsdal, C. ``Relativistic Thermodynamics, a Lagrangian Field Theory

for general flows including rotation"; arXiv 1106.2271 (revised 2015).

\noindent Gibbs, J.W., ``On the equilibrium of heterogeneous substances",

Trans.Conn.Acad. {\bf 3}, 108-248, 343-524 (1874-1878).

\noindent Goldstein, D., Handler, R. and Sirovich L., ``Modeling a No-Slip Flow Boundary with 

an external Force Field", J. Comp. Phys. {\bf 105} 354(1993).

\noindent Hall, H.E. and Vinen, W.F., ``The Rotation of Liquid Helium II. The Theory of

Mutual Friction in Uniformly Rotating Helium II", 

Proc. R. Soc. Lond. A 1956 238, 
doi: 10.1098/rspa.1956.0215 (1956).

\noindent Joseph, D.D. and   Renardy, Y., ``Couette flow of two fluids between
concentric cylinders", 

J . Fluid Mech. (1985). 150,  381-394 (1985).

\noindent Khalatnikov, I.M., {\it An Introduction to the Theory of Superfluidity},
Westview Press 2000. 

Original published as part of Frontiers in Physics Series, D.Pines (ed.) (1965).

\noindent Landau, L. D., and Lifschitz, E. M., Doklady Akad. NAUK {\bf 100} 669 (1955).

\noindent Landau, L. D., and Lifschitz, E. M., {\it Statistical Physics}, Pergamon Press, (1958). 





\noindent Lichnerowicz, A., {\it Th\'eories Relativistes de la Gravitation et de l'Electromagn\'etism},

 Masson et Cie, Editeurs, Paris (1955).

\noindent Lund,  F. and Reggee T.,  ``Unified Approach to strings and vortices with soliton 
 
solutions", Phys. Rev. D. {\bf 14} 1524-1548 (1976).

\noindent Martin, H., ``Reynolds, Maxwell and the radiometer, revisited",

Proc. 14'th Int. Heat Transfer Conf., August 08-13, 2010, Washington DC.

\noindent Minkowski 1908. For a modern discusssion and a list of references see Ridgely 1998.  

\noindent M\"uller, I., {\it A History of Thermodynamics}, Springer, Berlin (2007). 

\noindent Navier, L. M. Acad. Sci. {\bf 7}  375-394 (1827).

\noindent  Navier,C.L.M.H., ``M\'emoire sur les lois du mouvement des fluides", 

M\'em. Acad. Sci. Inst. France, {\bf 6}  389-440 (1882).

\noindent Ogievetskij, V.I. and Palubarinov, ``Minimal interactions between spin 0 and 

spin 1 fields",
J. Exptl. Theoret. Phys. (U.S.S.R.) {\bf 46} 1048-1055 (1964).

\noindent Onsager, L., see Eyink and Sreenivasen 2006. 

\noindent Priezjev, N.V.,    Darhuber, A.A. and  Troian, S.M., ``Slip length in sheared liquid films 

subject to mixed boundary conditions," Phys. Rev. E 71, 041608 (2005).

\noindent Prigogine, I., "Evolution criteria, variational properties and fluctuations", 

in 
{\it Thermodynamics, variational techniques and stability}, Donnelly et al, 

U. Chicago Press (1965).

\noindent Rasetti, M. and Regge, T., ``Vortices in HeII, current algebra and quantum 

knots" Physica {\bf 80A} 217-233 (1975).

\no Ridgely, C.T., ``Applying relativistic  electrodynamics to a rotating material medium", 

ajp-66-114-98 (1998).

\noindent Sedov, L.I., {\it A course in continuous mechanics}, Woolters-Noordhoff, Groeningen  (1971).





\noindent Schutz, B.F. Jr., ``Perfect fluids in General Relativity, Velocity potentials and a 

variational principle", Phys.Rev. {\bf D2}, 2762-2771 (1970).

\noindent Stanyukovich, K.P., {\it Unsteady motion of continuous media},

Pergamon Press N. Y. (1960).

\noindent Stokes, G.G., Trans. Cambridge Phil. Soc., {\bf 8}   287-319 (1843). 

\noindent Tolman, R.C., {\it Relativity, Thermodynamics and Cosmology},

Clarendon, Oxford (1934).

\noindent Weinberg, ``Gravitation and Cosmology: Principles and Applications of the 
General

Theory of Relativity", John Wiley, N.Y. (1972). 

\no Zheltukhin, A., ``On brane symmetries", arXiv.1409.6655.

\end{document}